\documentstyle[psfig]{caosp}
\begin{document}
\pubyear{1993}
\volume{23}
\firstpage{1} 
\hauthor{F. Leone}
\title{Helium in Chemically Peculiar Stars}
\author{F. Leone}
\institute{Catania Astrophysical Observatory} 


\maketitle

\begin{abstract} 
For the purpose of deriving the helium abundances in chemically
peculiar stars, the importance of assuming a correct helium
abundance has been investigated for determining the effective temperature and 
gravity of main sequence B-type stars, making full use of the present
capability of reproducing their helium lines.

Even if the flux distribution of main sequence B-type stars appears to
depend only on the effective temperature for any helium abundance, the
effective temperature, gravity and  helium abundance have to be determined
simultaneously by matching the Balmer line profiles.
New MULTI NLTE calculations, performed 
adopting ATLAS9 model atmospheres and updated helium atomic parameters, 
reproduce most of the observed equivalent widths of neutral helium lines
for main sequence B-type stars and they make us confident of the possibility
to correctly derive the helium abundance in chemically peculiar stars.

An application of previous methods to the helium rich star HD\,37017 shows
that helium could be stratified in the magnetic pole regions, as expected in
the framework of the diffusion theory in the presence of mass loss. 
\end{abstract}

\vspace{-3mm}
\section{Introduction}
\vspace{-1mm} 
Chemically peculiar (CP) stars can present helium lines whose strength is
 different
than in main sequence stars with equal Balmer lines and/or
flux distribution. Usually, helium lines are weaker in the coolest CP stars
and stronger in the hottest ones. 
As for other chemical elements, helium is supposed to be under- or overabundant
in the photosphere because of diffusion processes. Computations by
Vauclair et al. (1991) show that helium should also be stratified in the outer
stellar layers. 

All methods already known and used to determine stellar effective
temperatures and gravities, which are necessary to derive helium 
abundances, are calibrated assuming a solar chemical composition.
Thus, to correctly derive the peculiar helium abundance in CP stars, we
have to understand
\begin{itemize}
\item the importance of helium abundance in determining the effective
 temperature and gravity.
\end{itemize}

A further fundamental step to derive the helium abundance of CP stars is
the comprehension of
\begin{itemize}
\item the present capability of reproducing the helium lines of main
sequence stars.
\end{itemize}

\section{Importance of helium abundance in the determination of $T_{\rm eff}$
and $\log g$}
Because of the peculiar metal abundances, the flux distribution of CP stars is
different from the flux distribution of main sequence stars with the same
Balmer jump (Leckrone et al. 1974) and {\it ad hoc} methods are necessary to
determine the effective temperature of CP stars (Napiwotzki et al. 1993).

As to helium peculiar stars, Hauck \& North (1993) concluded that
{\it classical} photometric methods can be reliable to infer their
effective temperature.
To investigate the possibility that an inhomogeneous helium distribution
on the stellar surface could contribute to the observed photometric variability
of helium-weak stars, Catalano \& Leone (1996) computed the emergent flux
for ATLAS9 model atmospheres (Kurucz 1993) with solar helium abundance and
without helium. The negligible magnitude difference they found confirms the
result of Hauck \& North. 

To investigate the importance of a correct helium abundance assumption in
determining the effective temperature and gravity by matching Balmer lines,
Leone \& Manfr\`e (1997) have compared the H$_{\beta}$ line profile of some
helium-weak stars with SYNTHE spectra (Kurucz \& Avrett 1981) computed
adopting ATLAS9 model atmospheres with different helium 
abundances. These authors found that several models could match a single
observed line profile and concluded that
the effective temperature and gravity of helium-weak stars have to
be determined simultaneously with the helium abundances. The determination
of gravity appears to be very sensitive to the
helium abundance, probably because of its contribution to the electron
pressure, hence to the Stark effect.
As to the case of the helium weak star HD\,175362, Leone \& Manfr\`e found
that matching the H$_{\beta}$ line profile assuming a solar-composition 
atmosphere, the resulting effective temperature is underestimated by 1000 K
and the gravity by 0.25 dex.
\section{Present capability of reproducing helium lines of main sequence
B-type stars}
Leone \& Lanzafame (1998) found that the NLTE calculations which are
available in the literature are not reliable in deriving the helium abundance
since they are not able to match the observed equivalent width of neutral
helium lines of main sequence stars with 10000 $< T_{\rm eff} <$ 30000 K.
These authors performed
new MULTI (Carllson 1986) NLTE calculations combining line blanketing ATLAS9
model atmospheres and the latest helium atomic data from the NIST database.
The agreement between theory and observations found for most of the 
neutral helium
lines of main sequence stars reinforces our confidence in the possibility to
correctly derive the helium abundance in chemically peculiar stars.  
\section{Evidence of possible helium stratification in the photospheres of CP
stars: the case of HD\,37017}
Calculations by Vauclair et al. (1991) show that helium should appear as
normal or slightly underabundant at the magnetic equator of the hottest
magnetic stars, according to the strength of the horizontal magnetic field.
In the presence of mass-loss, helium accumulates at the magnetic poles. 
Moreover, according to these calculations, helium is stratified in the 
atmospheres of magnetic CP stars: helium abundance increases with optical
depth, reaches a maximum and then decreases.
The position of the helium abundance maximum depends on the effective
temperature, mass loss and diffusion strength.

\begin{figure}[htbp]
\centerline{\hbox{ 
\psfig{file=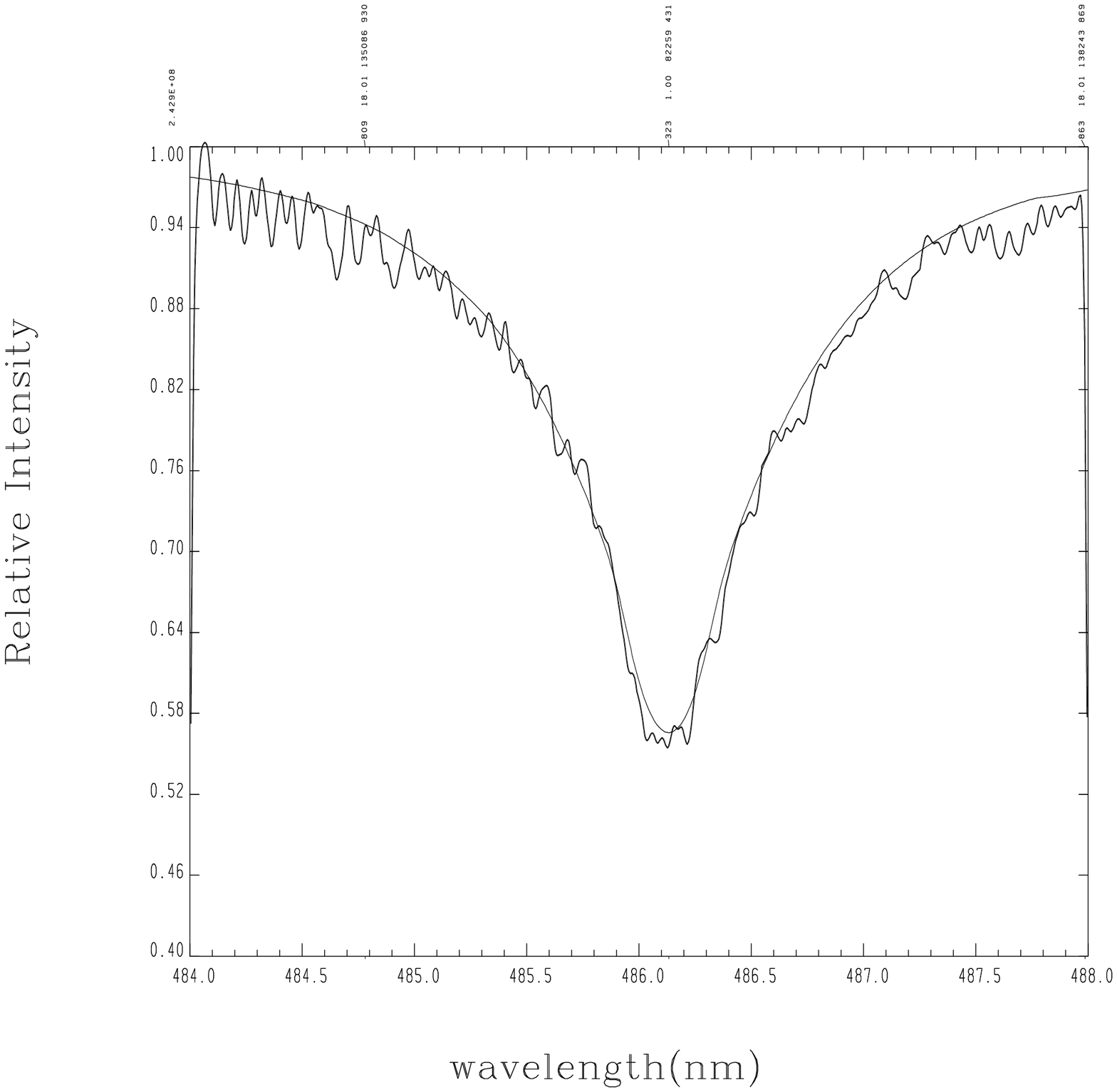,width=6.2cm}
\psfig{file=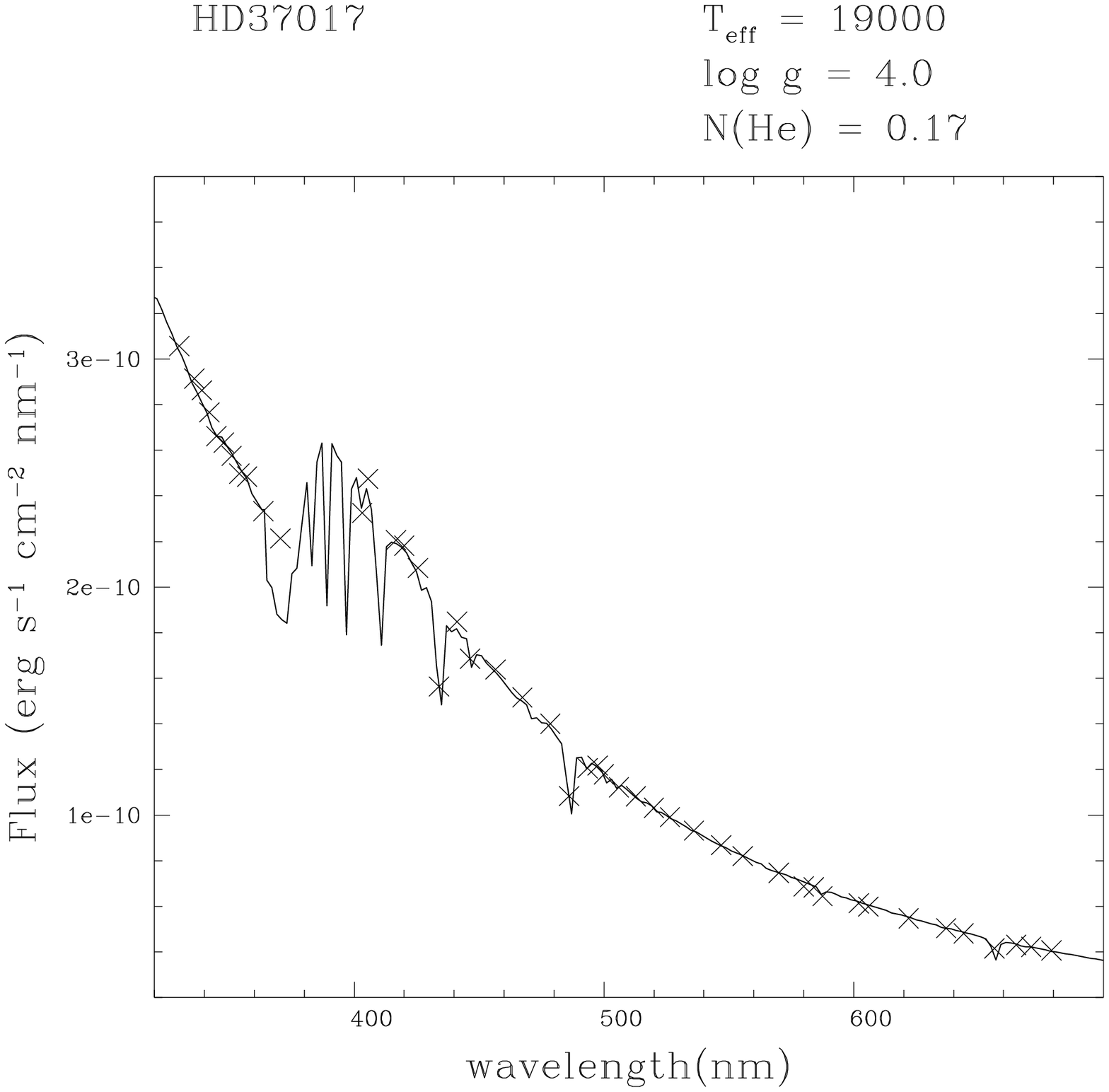,width=6.2cm}
           }     }
\caption{The H$_{\beta}$ line profile and visible flux distribution of
HD\,37017 are matched assuming $T_{\rm eff}$ = 19000 K, log\,g = 4.0 and 
n(He)/n(H) = 0.17, using an ATLAS9 model atmosphere.}
\end{figure}

To verify the validity of the results by Vauclair et al. (1991), 
spectroscopic observations have been carried out
of the helium lines (in the range 410 - 710 nm) of 
the helium-rich star HD\,37017 during the phases of minimum and maximum
helium line strengths. According to Bohlender et al. (1987), at these phases
the line of sight lies in the magnetic equatorial plane and it is close to
the negative magnetic pole respectively.  

To look for evidence of helium stratification in the atmosphere of
HD\,37017, the helium lines which are mainly formed in different
photospheric layers and which are reliably reproduced for main sequence stars
by calculations of Leone \& Lanzafame have been selected.  

By using the iterative procedure described by Leone \& Manfr\`e (1997),
with the difference that helium abundances are derived here by performing 
NLTE calculations according to Leone \& Lanzafame (1998), the effective 
temperature and gravity of HD\,37017 have been
determined simultaneously with the helium abundance
by matching the H$_{\beta}$ line profile at the phase of helium line strength 
minimum (Fig.\,1).
As expected, the emergent flux of the corresponding ATLAS9 model atmosphere
well represents also the flux
distribution observed by Adelman \& Pyper (1985) (Fig.\,1)

When we are looking at the magnetic
equator, all helium abundance values almost coincide, while they are derived 
from lines which are mainly formed in different photospheric layers,
as expected for a non-stratified atmosphere.
When we are looking at the negative
magnetic pole, we obtain abundance values which decrease with optical
depth (Fig.\,2). This decrement cannot be removed by changing the 
microturbulence velocity, the effective temperature and/or gravity. 

Thus, the results of Vauclair and co-workers appear to be confirmed: helium is 
more abundant at the magnetic poles than at the magnetic equator, and it is 
stratified in the outer photospheric layers of the hottest CP stars.

\begin{figure}[htbp]
\centerline{\hbox{ 
\psfig{file=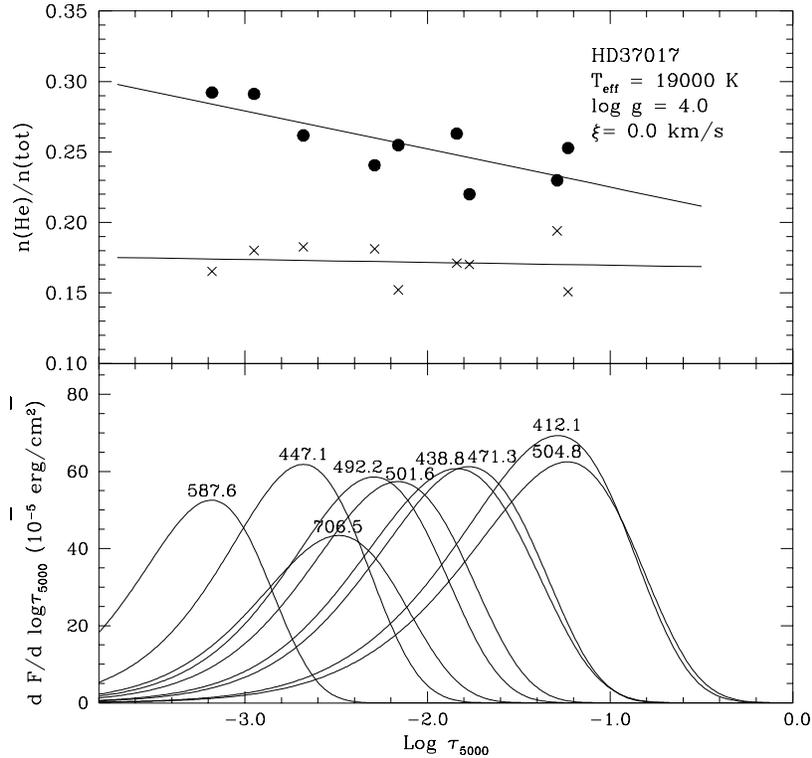,width=11cm}
                 }} 
\caption{In the top panel, abundances are reported as a function of the optical
depth for the helium rich star HD\,37017.
Crosses represents the abundances derived at the phase of minimum strength of
helium lines. Dots correspond to the maximum strength. In the bottom panel
are shown the source functions for the line cores.}
\end{figure}

\section{Conclusion}
Since the stellar flux distribution of B-type stars depends only slightly
on gravity and helium abundance, it can be used to determine the effective
temperature of chemically peculiar stars. Anyway, since the Balmer lines
are necessary to infer the gravity and are influenced by helium abundance,
the effective temperature, gravity and helium abundance have to be determined 
simultaneously by matching their profiles. 

The NLTE calculations which are available in the literature do not reproduce
the observed equivalent width of the neutral helium lines for main sequence
B-type stars. Before deriving the helium abundance, the reliability of the
method used should be at least tested on main sequence B-type stars before
it is applied to chemically peculiar stars. 

Using the method described by Leone \& Manfr\`e (1997) to determine the
stellar parameters from the Balmer line profiles, and using the NLTE 
calculations by Leone \& Lanzafame (1998) of neutral helium lines,
the helium abundance
has been derived for the helium-rich star HD\,37017 from a sample of lines
which are mainly formed in different atmospheric layers. It appears that
helium abundance decreases with optical depth in the magnetic pole regions 
while it remains constant with optical depth in the magnetic equatorial region.
This result is consistent with the calculations by Vauclair et al. (1991),
which foresee overabundant and stratified helium in the atmospheric regions
close to the magnetic poles.

\end{document}